\documentclass[twocolumn,pra,aps,superscriptaddress,showpacs]{revtex4}
\usepackage{mathrsfs}
\usepackage{graphicx}
\begin{document}
\title {Resonantly enhanced coherence by laser-assisted tunneling}
\author{Hong-Xia Hao}
\author{Shiping Feng}
\affiliation{Department of Physics, Beijing Normal University, Beijing 100875, China}
\author{Shi-Jie Yang\footnote{Corresponding author: yangshijie@tsinghua.org.cn}}
\affiliation{Department of Physics, Beijing Normal University, Beijing 100875, China}
\affiliation{State Key Laboratory of Theoretical Physics, Institute of Theoretical Physics, Chinese
Academy of Sciences, Beijing 100190, China}
\begin{abstract}
We study quantum coherence of strongly interacting cold bosons in a double-well potential driven by a laser field. The system is initially in a Fock state and, for either with or without a static tilting field, evolves into the coherent states. The coherence is resonantly enhanced by the photon-assisted tunneling. For the tilted wells, it reveals a two-branch pattern which corresponds to the multiple photon absorption or emission, respectively.

\end{abstract}
\maketitle

\section{introduction}
Ultracold atoms confined in optical lattices are well suited to explore the many-body phenomena in the condensed matter physics. Experimental and theoretical investigations regarding this field have undergone amazing  progress. The phenomena of quantum tunneling\cite{Winkler,Folling,Zollner}, the superfluid-Mott phase transition\cite{Kuklov,Yang,Yang1}, and the disorder effects\cite{Lye,Fallani,Zhang}, etc., were explored extensively. In a double-well, atomic Josephson oscillations take place in a Bose-Einstein condensate (BEC) when the initial population imbalance is below a critical value\cite{Jack,Zapata}, and a phenomenon of macroscopic quantum self-trapping where the atoms essentially stay in one well, is observed in experiments\cite{Anker,Albiez}.

In recent years, the non-equilibrium phenomena in cold atom systems have attracted great interests\cite{Yukalov2}. The behavior of the collapse and revival of the matter wave field is demonstrated in the dynamical evolution of the interference pattern\cite{Greiner}. Researches on periodic shaking lattices have invoked effects ranging from coherent destruction of tunneling, dynamic localization, to field-induced barrier transparency, super Bloch oscillations, phase-jumps and dynamics of bound pairs as well as artificial magnetic fields in the many body physics\cite{Kayanuma1,Eckardt1,Lignier,Grossmann,Zenesini,Weiss,Longhi,Sudheesh,Haller,Ridinger,Weiss1}.
One of the remarkable effect of the periodically driving potential is the photon-assisted tunneling. The "photons" are time-dependent potential modulations in the kilohertz regime rather than real
photons. This effect may induce a tunable superfluid-Mott insulator transition as the external driving
frequency matches the interaction energy\cite{Eckardt2} and has been observed experimentally\cite{Sias}.

In this paper we study the quantum dynamics of $N$ bosons confined in a double-well under the influence of a laser field. By making a two mode approximation, the Hamiltonian is written as\cite{Yukalov}
\begin{eqnarray}
\hat{H}(t)&=&-\upsilon(\hat{a}_1^\dag \hat{a}_2+\hat{a}_2^\dag \hat{a}_1)+\frac{1}{2}U\sum_{i=1,2}\hat{n}_i(\hat{n}_i-1)\nonumber\\
& &+\frac{1}{2}(\Delta+A\cos\Omega
t)(\hat{n}_1-\hat{n}_2),\label{hamt}
\end{eqnarray}
where $\hat{a}_{1(2)}^\dag$ and $\hat{a}_{2(1)}$ respectively denote the creation and annihilation operators for bosonic atoms localized in either side of the well. $\upsilon$ is the hopping matrix element between the neighboring sites. $\hat{n}_i$ is the number operator at the $i$th site. $U$ is the Hubbard interaction between a pair of bosons occupying the same site, which can be adjusted by the Feshbach resonance technique. $\Delta$ is the static energy difference of the tilted wells. $A$ and $\Omega$ are respectively the amplitude and frequency of the driving laser field.

We focus on the strong interaction regimes $U\gg\upsilon$, either with ($\Delta\neq0$) or without ($\Delta=0$) a static tilting field applying on the double-well. We investigate, with the help of both analytical arguments and numerical simulations, the dependence of the coherence on the driving field as the system evolving from the initial Fock state. As the driving frequency matches the sum of the interaction energy and the tilted energy difference, the system resonantly absorbs or emits energy
that exhibits a multiple photon process. The single particle tunneling is strengthened and the quantum coherence are greatly enhanced.

This paper is organized as follows. In Sec. II, we give a theoretical description of the resonant coupling of the many-boson system with the driving field. In Sce.III, we numerically display, by introducing a physical quantity to depict coherence degree, the temporal evolution of the system for the untilted wells. Section IV presents the coherence evolution for the tilted wells. A brief summary is included in Sec. V.

\section{theoretical description}
We divide the time-dependent Hamiltonian (\ref{hamt}) into two parts, the hopping part $H_1$ and the interaction $H_0(t)$,
\begin{eqnarray}
\hat{H}_1=-\upsilon(\hat{a}_1^\dag \hat{a}_2+\hat{a}_2^\dag
\hat{a}_1),\label{ham1}
\end{eqnarray}
\begin{eqnarray}
\hat{H}_0(t)=\frac{1}{2}U\sum_{i=1,2}\hat{n}_i(\hat{n}_i-1)+\frac{1}{2}(\Delta+A\cos\Omega
t)(\hat{n}_1-\hat{n}_2).\label{ham0t}
\end{eqnarray}
The operators (\ref{ham1}) and (\ref{ham0t}) are represented with the complete set of Fock bases \{$|N,0\rangle, |N-1,1\rangle, \cdots,|0,N\rangle $\} as
\begin{eqnarray}
\langle
k|\hat{H}_1|j\rangle=-(\delta_{k,j-1}+\delta_{k,j+1})\upsilon\sqrt{N-k}\sqrt{k+1},
\end{eqnarray}
\begin{eqnarray}
\langle k|\hat{H}_0(t)|j\rangle &=&\delta_{k,j}[\frac{U}{2}(N^2-2Nk-N+2k^2)\nonumber\\
& &+\frac{1}{2}(\Delta+A\cos\Omega t)(N-2k),
\end{eqnarray}
where $|k\rangle=|k,N-k\rangle$ with $k=0,\cdots,N$. The many-body state is expressed as
\begin{eqnarray}
|\psi(t)\rangle=\sum_{k=0}^N c_{k}(t)|N-k,k\rangle.
\end{eqnarray}
By making the transformation to the interaction picture through\cite{Weiss2,Weiss3}
\begin{eqnarray}
c_k(t)=a_k(t)e^{-i\int_0^t\langle k|\hat{H}_0(t')|k\rangle dt'},
\end{eqnarray}
and inserting the state into the Heisenberg equation of motion, one obtains
\begin{eqnarray}
i\dot{a}_k(t)&=&\langle k|\hat{H}_1|k+1\rangle h_{k+1}(t)a_{k+1}(t)\nonumber\\
& &+\langle k|\hat{H}_1|k-1\rangle h_{k-1}(t)a_{k-1}(t),\label{motion}
\end{eqnarray}
where the phase factor reads
\begin{eqnarray}
h_{k+1}(t)&=&e^{-i(U+\Delta) t-\frac{iA}{\Omega}\sin\Omega t},\nonumber\\
h_{k-1}(t)&=&e^{-i(U-\Delta) t-\frac{iA}{\Omega}\sin\Omega t},
\end{eqnarray}
for the initial Fock state $|\psi(0)\rangle=|N/2,N/2\rangle$. The phase factor can be computed by employing the expansion $e^{iz\sin(\Omega t)}=\sum_{n=-\infty}^{\infty}J_n(z)e^{in\Omega t}$, where $J_n(z)$ is the $n$th-order Bessel's function of the first kind, yielding
\begin{eqnarray}
h_{k+1}(t)&=&\sum_{n=-\infty}^{\infty}J_n(A/\Omega)e^{i(n\Omega-U-\Delta)t},\nonumber\\
h_{k-1}(t)&=&\sum_{n=-\infty}^{\infty}J_n(A/\Omega)e^{i(n\Omega-U+\Delta)t}.\label{phase}
\end{eqnarray}

In the limit of strong interactions $U/\upsilon\gg 1$, the phase factor $h_k(t)$ contains rapidly oscillating terms which are averaged over the time period $T=2\pi/\Omega$. With
\begin{eqnarray}
\frac{1}{T}\int_{0}^{T}e^{i(n\Omega-U)t}=\delta(n\Omega-U),
\end{eqnarray}
it follows that
\begin{eqnarray}
h_{k+1}&=&\sum_{n=-\infty}^{\infty}J_n(A/\Omega)\delta(n\Omega-U-\Delta),\nonumber\\
h_{k-1}&=&\sum_{n=-\infty}^{\infty}J_n(A/\Omega)\delta(n\Omega-U+\Delta).
\end{eqnarray}
As indicated in Eq.(\ref{motion}), the lower frequency parts of the motion are contributed by the hopping terms which can be taken as quasi-static when we take the time-average over the high frequency parts.

The resonance occurs at
\begin{equation}
n\Omega=U\pm\Delta.\label{resonance}
\end{equation}
Under this circumstance, the equation of motion (\ref{motion}) reduce to
\begin{eqnarray}
i\dot{a}_k(t)&=&\langle k|J_n(A/\Omega)\hat{H}_1|k+1\rangle a_{k+1}(t)\nonumber\\
& &+\langle k|J_n(A/\Omega)\hat{H}_1|k-1\rangle a_{k-1}(t),\label{motion1}
\end{eqnarray}
which constitute a series of coupled equations of free oscillators with renormalized tunneling parameter as
\begin{equation}
\widetilde{\upsilon}_n=\upsilon J_n(A/\Omega).\label{eff2}
\end{equation}

\section{untilted wells}
We give the exact results by numerically solving the Heisenberg equation of motion based on the Hamiltonian (\ref{hamt}). To characterize the coherence of the system, we adopt a quantity of coherence degree introduced by one of the authors\cite{Yang},
\begin{equation}
\alpha(t)=\frac{|\lambda_1-\lambda_2 |}{\lambda_1+\lambda_2},
\end{equation}
where $\lambda_1$ and $\lambda_2$ are the eigenvalues of the single-particle density $\rho_{\mu\nu}(t)=\langle \psi(t)|\hat{a}_\mu^\dag \hat{a}_\nu |\psi(t)\rangle$ ($\mu,\nu=1,2$)\cite{Penrose,Mueller}. As $\alpha(t)\rightarrow 1$, the system is in the coherent (quasi-coherent) state since in this case there is only one large eigenvalue of matrix $\rho_{\mu\nu}$. Whereas $\alpha(t)\rightarrow 0$ indicates the system is in the decoherent or fragmented state because there are two densely populated natural orbits. In the weak interaction or strong tunneling
limit ($U/\upsilon\ll 1$), each atom is in a coherent superposition of the left-well and right-well states. In the strong interaction or weak tunneling limit ($U/\upsilon\gg 1$), the tunneling term is
negligible. The state is a product of the number operators for the left and right wells. This regime is analogous to the Mott insulator (MI) phase in optical lattices.

\begin{figure}
\begin{center}
\includegraphics*[width=9cm]{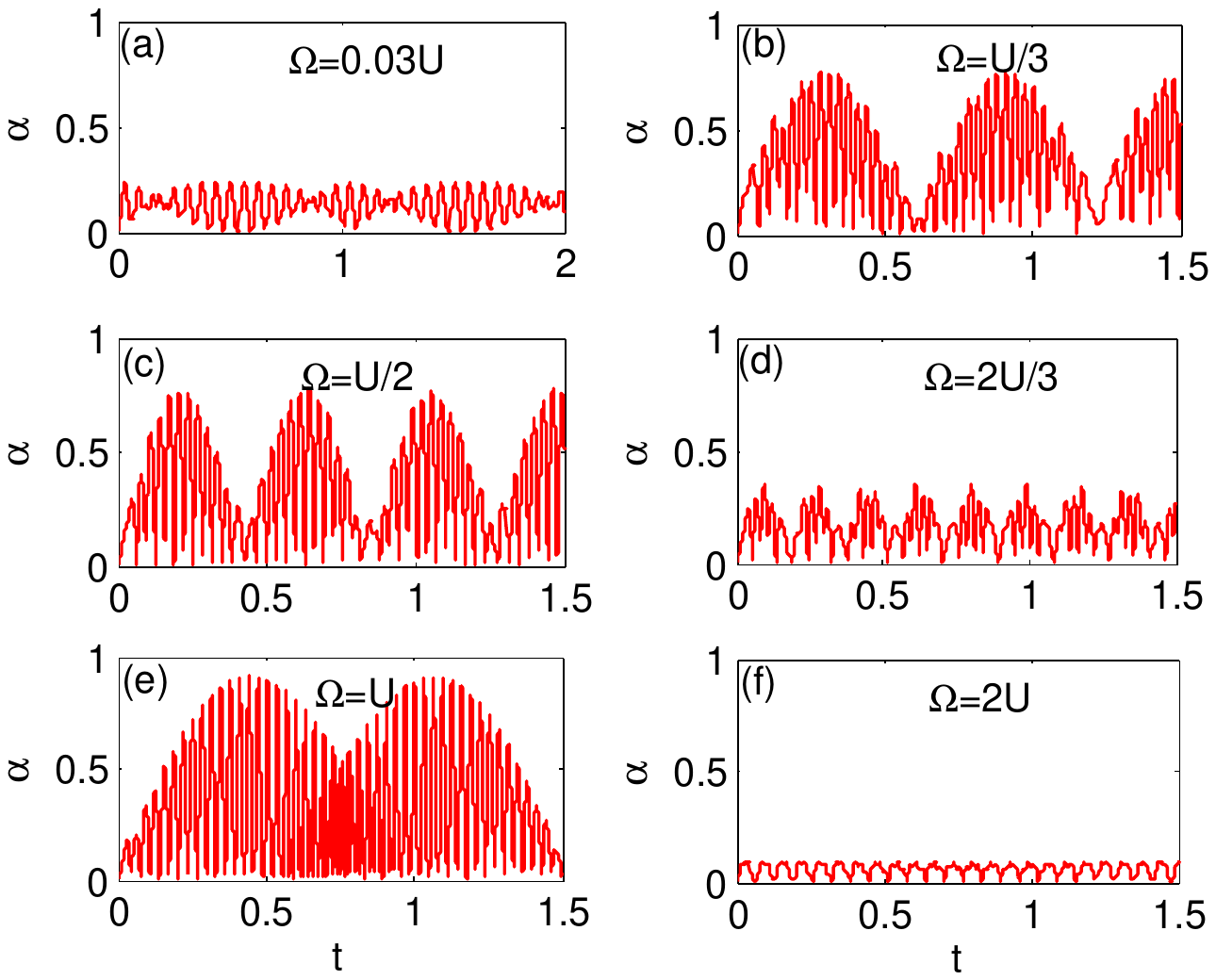}
\caption{(Color online) Temporal evolution of $\alpha(t)$ in the untilted double-well driving by a periodical field at fixed $A/\Omega=3.2$. (a-f) $\Omega=0.03U, U/3, U/2, 2U/3, U, 2U$, respectively. The photon-assisted tunneling occurs as $\Omega$ is a fraction of $U$, as shown in (b,c,e).}
\end{center}
\end{figure}

\begin{figure}
\begin{center}
\includegraphics*[width=9cm]{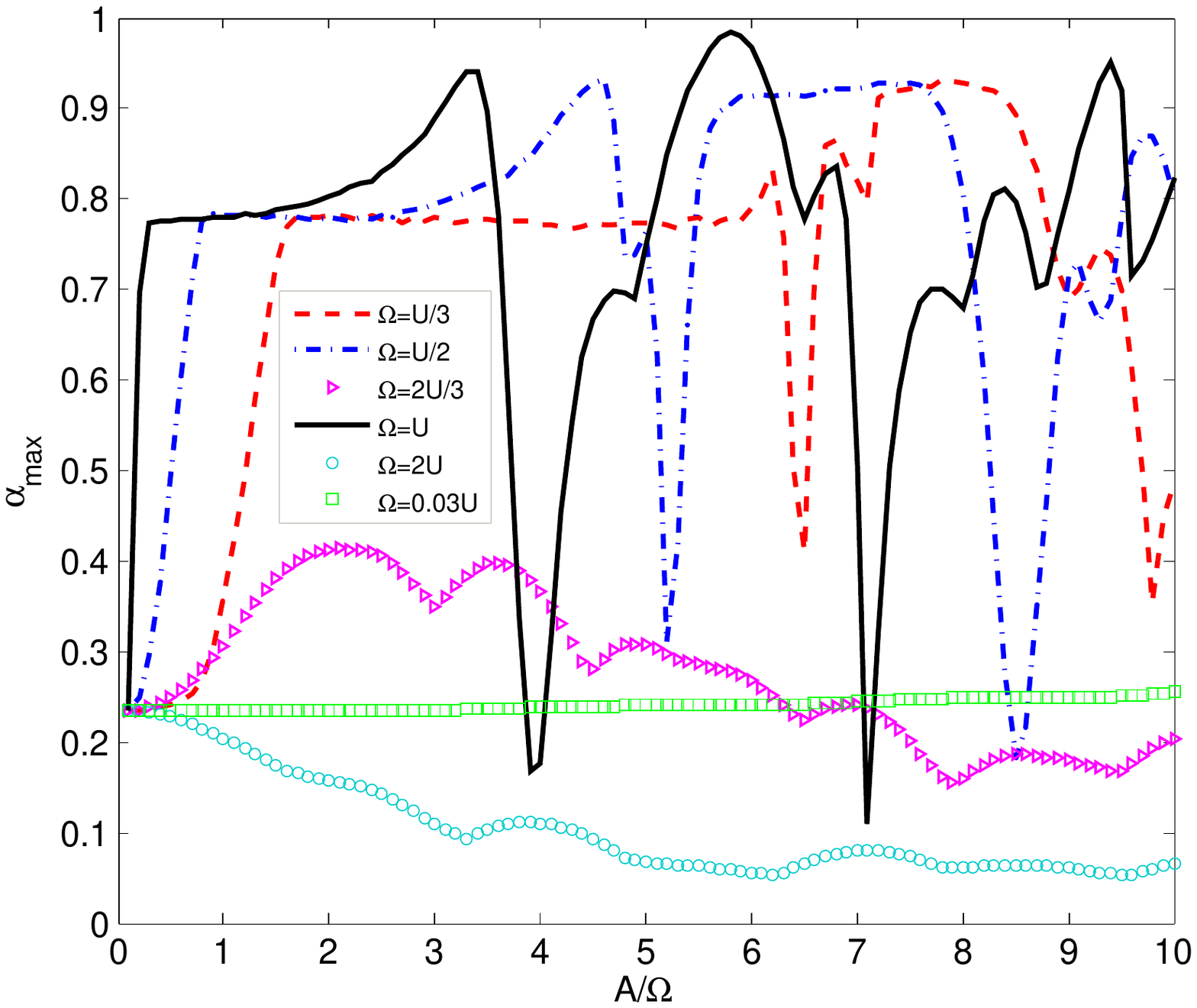}
\caption{(Color online) The maximum coherence $\alpha_\textrm{max}$ versus the ratio $A/\Omega$ in the untilted double-well for various driving frequencies. Resonant tunneling takes place at frequencies $\Omega=U/3,U/2,U$, with some dips which come from zeros of the renormalized tunneling matrix elements.}
\end{center}
\end{figure}

In this section we consider the untilted double-well case ($\Delta=0$). We take the units of $\upsilon=1$ and fix the Hubbard energy $U=100$ and the particle number $N=10$ as example. The
initial state is a symmetric Fock state $|\Psi(0)\rangle=|N/2,N/2\rangle$. In absence of the driving field ($A=0$), the single particle tunneling is damped and the coherence is suppressed. The driving field may assist the particle tunneling between the wells through exchanging energy with the bosons and enhance the coherence. Figure 1 shows the real-time evolution of the coherence degree $\alpha$ for various driving frequencies at a fixed ratio $A/\Omega=3.2$. In Fig.1(a,d,e) where $\Omega=0.03U, 2U/3, 2U$, respectively, the coherence $\alpha(t)$ remains very low which indicates that the single particle tunneling is still suppressed. The bosons can nearly exchange the energy with the driving field. In contrast, in Fig.1(b,c,f) which respectively corresponds to the driving frequencies $\Omega=U/3, U/2, U$, the coherence are greatly increased in a simple oscillating mode which implies
single particle tunneling is strengthened through exchanging energy with the driving field. This photon-assisted tunneling occurs resonantly as the driving frequency satisfies $\Omega=U/n$ (with $n$ integer), in analogy to the $n$-photon absorption process.

\begin{figure}[t]
\begin{center}
\includegraphics*[width=9cm]{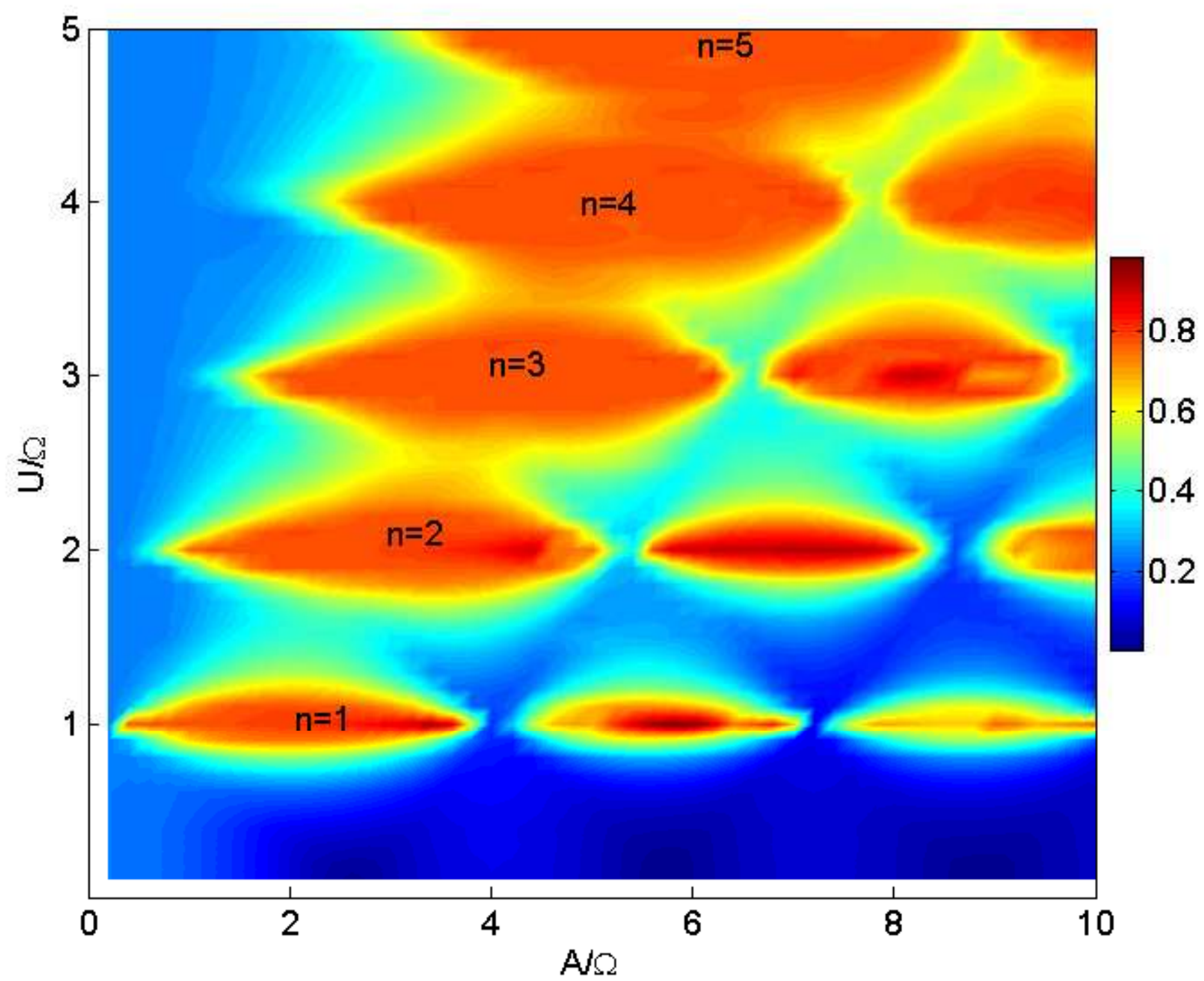}
\caption{(Color online) Topographical map of $\alpha_\textrm{max}$ versus $A/\Omega$ and $U/\Omega$ in the untilted double-well. The horizonal ridges indicate resonant enhancement of the coherence. The integer numbers specify the orders of the resonant absorption of multiple photons.}
\end{center}
\end{figure}

In order to give a more explicit description, we investigate the maximum value of the coherence degree $\alpha(t)$. Figure 2 illustrates $\alpha_\textrm{max}$ versus the ratio $A/\Omega$ for different driving frequencies. The three lower curves are for $\Omega=0.03U, 2U, 2U/3$, respectively. In comparison, the coherence is greatly enhanced for frequencies $\Omega=U/3, U/2, U$, except some dips at special values of ratio $A/\Omega$ which are the zeros of the $n$th Bessel's functions $J_n(A/\Omega)$. As explained in Sec.II, under the external driving field the effective tunneling matrix element corresponding to the $n$th order resonant absorption takes the renormalized form of Eq.(\ref{eff2}). It causes destructive tunneling as the augment equals to the zeros of the corresponding Bessel's functions. Our numerical results agree with the prediction quite well.

Figure 3 displays the topographical graph of the $\alpha_\textrm{max}$ versus $U/\Omega$ and $A/\Omega$. The horizonal ridges indicated by the integer numbers ($n=1,2,3,4,5$) clearly reveal the resonant tunneling which are assisted by the multiple photon absorption. The resonances happen as the driving frequency is an integer fraction of the interaction strength, $U/\Omega=n$, as claimed in Eq.(\ref{resonance}). In this process the system absorbs $n$-photons. The ridge is broken at dips which correspond to the correlated destructive tunneling as $A/\Omega$ take the zeros of the Bessel's functions.

\begin{figure}
\begin{center}
\includegraphics*[width=9cm]{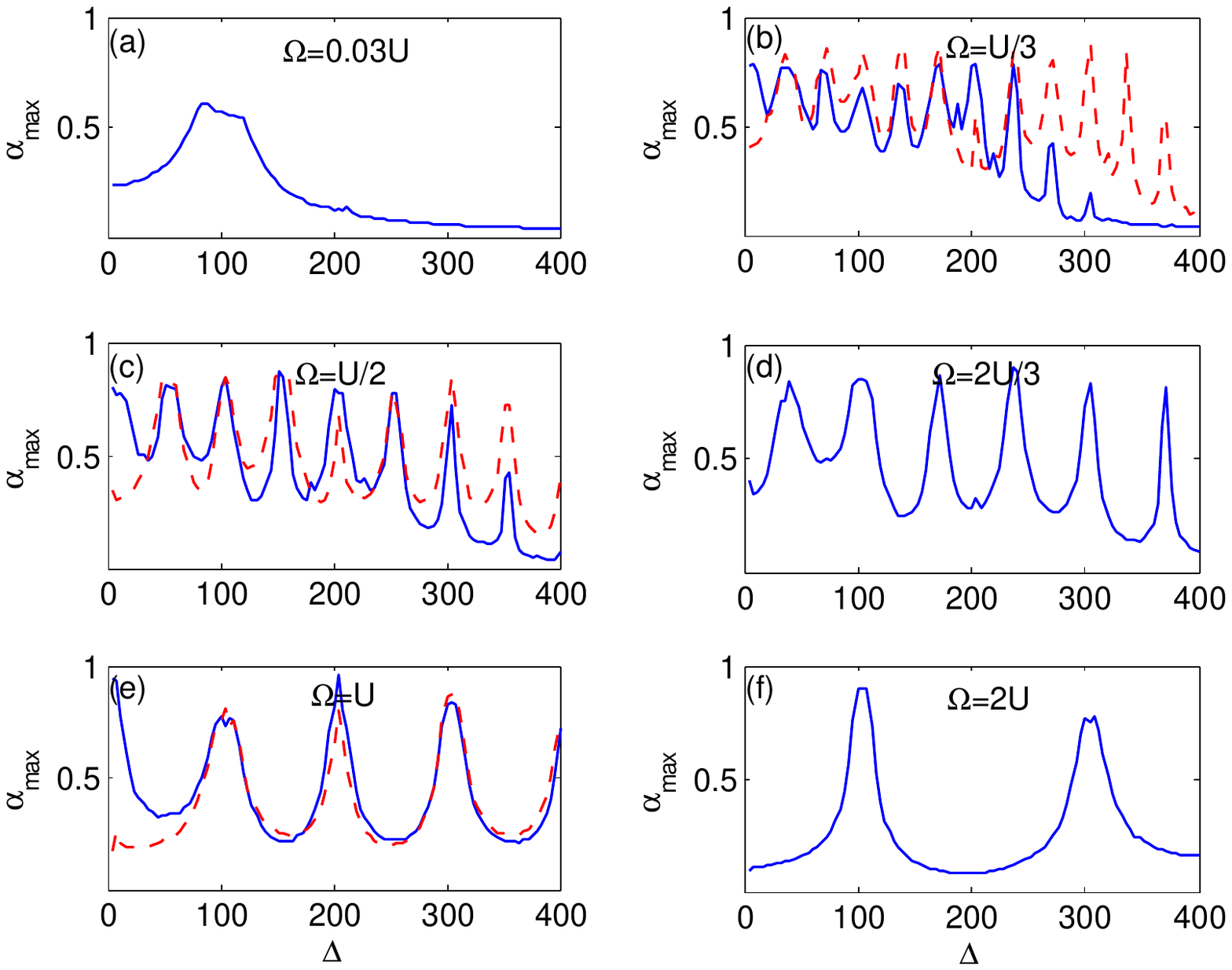}
\caption{(Color online) Resonant enhancement of coherence versus the static tilting $\Delta$ for various driving frequencies. (a-f) $\Omega=0.03U, U/3, U/2, 2U/3, U, 2U$, respectively. The solid curves are calculated for $A/\Omega=3.2$. The dashed curves in (b), (c) and (e) are calculated respectively for $A/\Omega=3.8$, $A/\Omega=5.1$, and $A/\Omega=6.3$, which are zeros of the Bessel's functions $J_1(z)$, $J_2(z)$ and $J_3(z)$. In these cases, the first resonant peak is damped by the destructive coherent tunneling.}
\end{center}
\end{figure}

\section{tilted wells}
It is more intriguing as the system is tilted which induces a finite potential energy difference ($\Delta\neq 0$) between the two wells. This static tilting can be realized experimentally by a constant acceleration or a gradient magnetic field. The evolution from the initial state $|\Psi(0)\rangle=|N/2,N/2\rangle$ to $|N/2+1,N/2-1\rangle$ or $|N/2-1,N/2+1\rangle$ through one-particle hopping induce an energy difference of either $U+\Delta$ or $U-\Delta$, which takes place as it matches the multiple photon absorption or emission energy. The competition between the Hubbard energy and the tilting potential leads to reducing or raising of the effective particle tunneling\cite{Creffield}. The additional driving field renormalizes the particle tunneling under specific circumstance.

\begin{figure}[t]
\begin{center}
\includegraphics*[width=9cm]{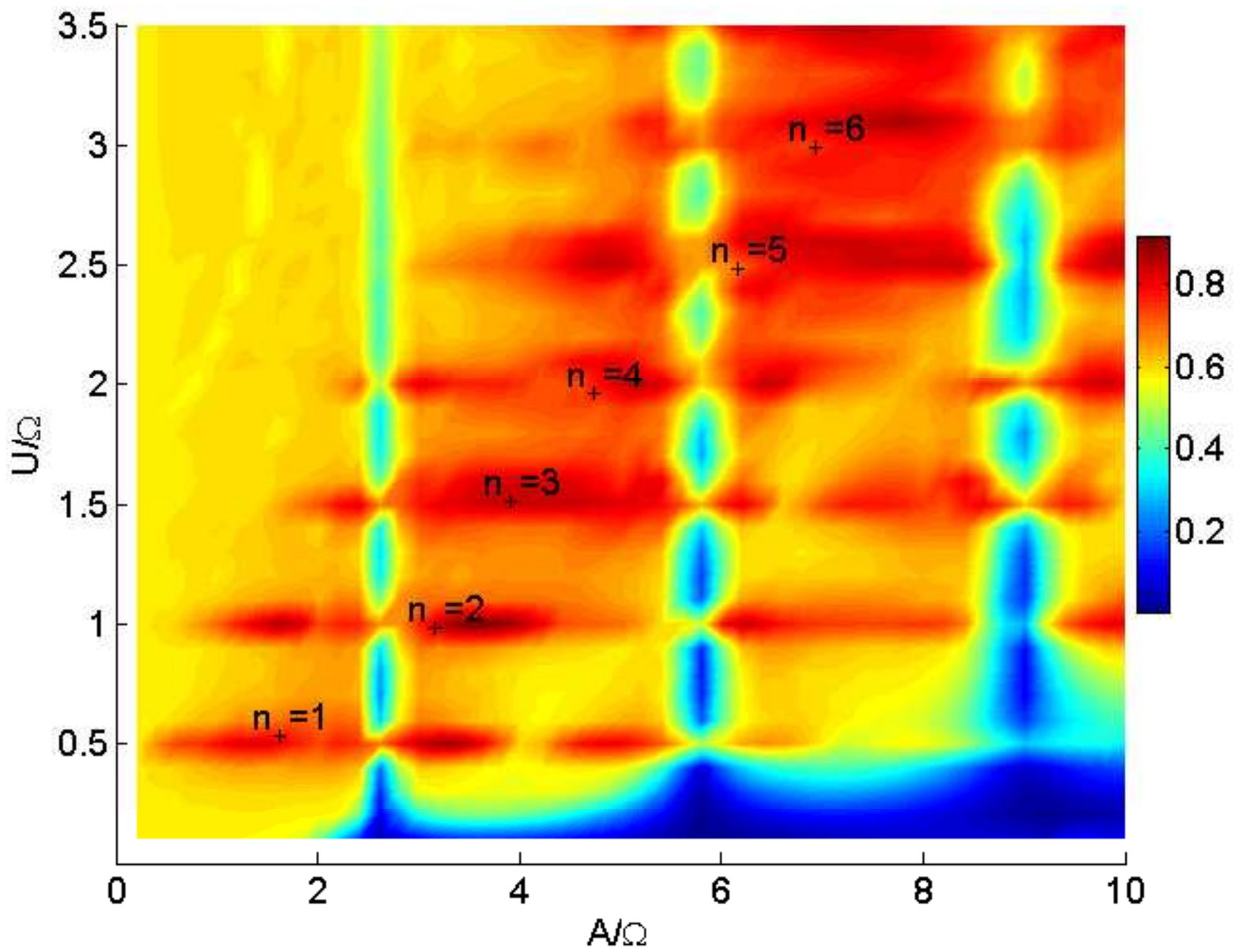}
\caption{(Color online) The same as in Fig.3 for the tilted wells $\Delta=U$. A striking feature is the three vertical valleys which correspond to the zeros of the zero-th order of Bessel's functions $J_0(z)$ where the coherent tunneling is destructive. The valleys hunch up as they intersect with the resonant ridges.}
\end{center}
\end{figure}

Figure 4 display the numerical results of $\alpha_\textrm{max}$ versus the static tilting $\Delta$ at fixed $A/\Omega=3.2$. From (a) to (f), the driving frequency takes $\Omega=0.03U, U/3, U/2, 2U/3, U, 2U$, respectively. Evidently, the equidistant resonant peaks take place as the parameters satisfy the following relation,
\begin{equation}
n_{\pm}\Omega=U\pm\Delta,\label{peak}
\end{equation}
where $n_{\pm}$ are integers which respectively indicate two coinciding sets of resonant peaks as $\Delta\neq 0$. The $\pm$ sign on the right-hand side of (\ref{peak}) respectively corresponds to a particle hop from the lower well to the higher well and vice versa. A special case happens at the untilting point $\Delta=0$, where only one set of resonant peak occurs. This peak can be damped by take the values of $A/\Omega$ as the zeros of the corresponding Bessel's functions, as shown in Fig.4(b,c,e) where the dashed curves correspond respectively to the 3rd, 2nd and 1st order of the Bessel's functions. All other peaks for $\Delta\neq 0$ are robust regardless of the values of the driving parameters.

\begin{figure}
\begin{center}
\includegraphics*[width=9cm]{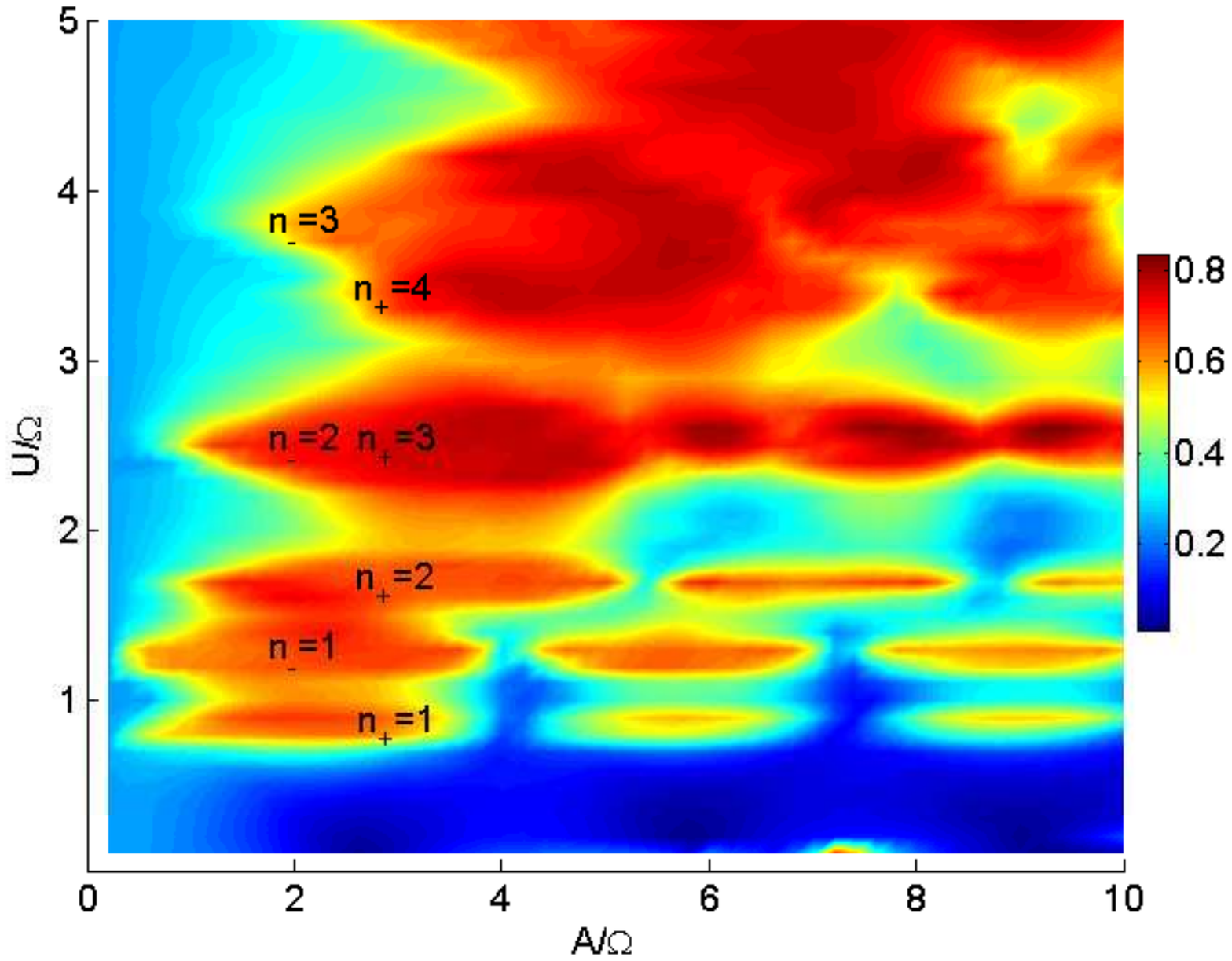}
\caption{(Color online) The same as in Fig.5 for tilting $\Delta=U/5$. One observes that the vertical valleys are absent while two sets of horizonal ridges which are indicated by integer numbers $n_+$ and $n_-$ are present.}
\end{center}
\end{figure}

Figure 5 shows the $\alpha_\textrm{max}$ for tilting $\Delta=U$. The horizontal ridges indicated by $n_+=1,2,3,\cdots$ correspond to the resonant tunneling to the state $|N/2+1,N/2-1\rangle$. In comparison to Fig.3, the multiple photon-assisted enhancement are observed at $U/\Omega=0.5, 1.5, 2.5,\cdots$, which satisfy the relation of $n_+\Omega=U+\Delta$. The $n_-$ series are absent because $U-\Delta=0$. The vertical valleys are from the zero points of the zero-th order of Bessel's functions, $A/\Omega=2.4, 5.52, 8.6$, etc. Fig.6 uses the same parameters but with $\Delta=U/5$ as in Fig.5. In this case, two branches of resonant enhancement $n_+$ and $n_-$ are clearly present. The ridges for $n_+=n_-$ have the same dips due to they belong to the same order of Bessel's functions and have the same effective tunneling parameter. Besides, the ridges of $n_-=2$ and $n_+=3$ coincide because $(U-\Delta)/2=(U+\Delta)/3$.

Finally, in Fig.7 we illustrate $\alpha_\textrm{max}$ dependence on the ratios of $\Delta/U$ and $\Omega/U$ in the tilted double-well system by fixing $A/\Omega=3.2$. The resonant tunneling to the state $|N/2+1,N/2-1\rangle$ exhibits straight ridges with positive slope which are marked in the figure with $n_+=1, 2, 3, 4, \cdots$. For the other branch $n_-\Omega=U-\Delta$, the resonant enhancement displays a symmetrical fan-shape structure start from $n_-=0$ ($\Delta/U$=1) at the center to $n_-=\pm1,\pm2,\cdots$ successively. The ridges with positive value of $n_-$ lie on the left-side of the fan which indicate absorption of photons, while the ridges with negative value of $n_-$ lie on the right-side of the fan which indicate emission of photons. The numerical results agree with the theoretical analysis very well.

\begin{figure}[t]
\begin{center}
\includegraphics*[width=9cm]{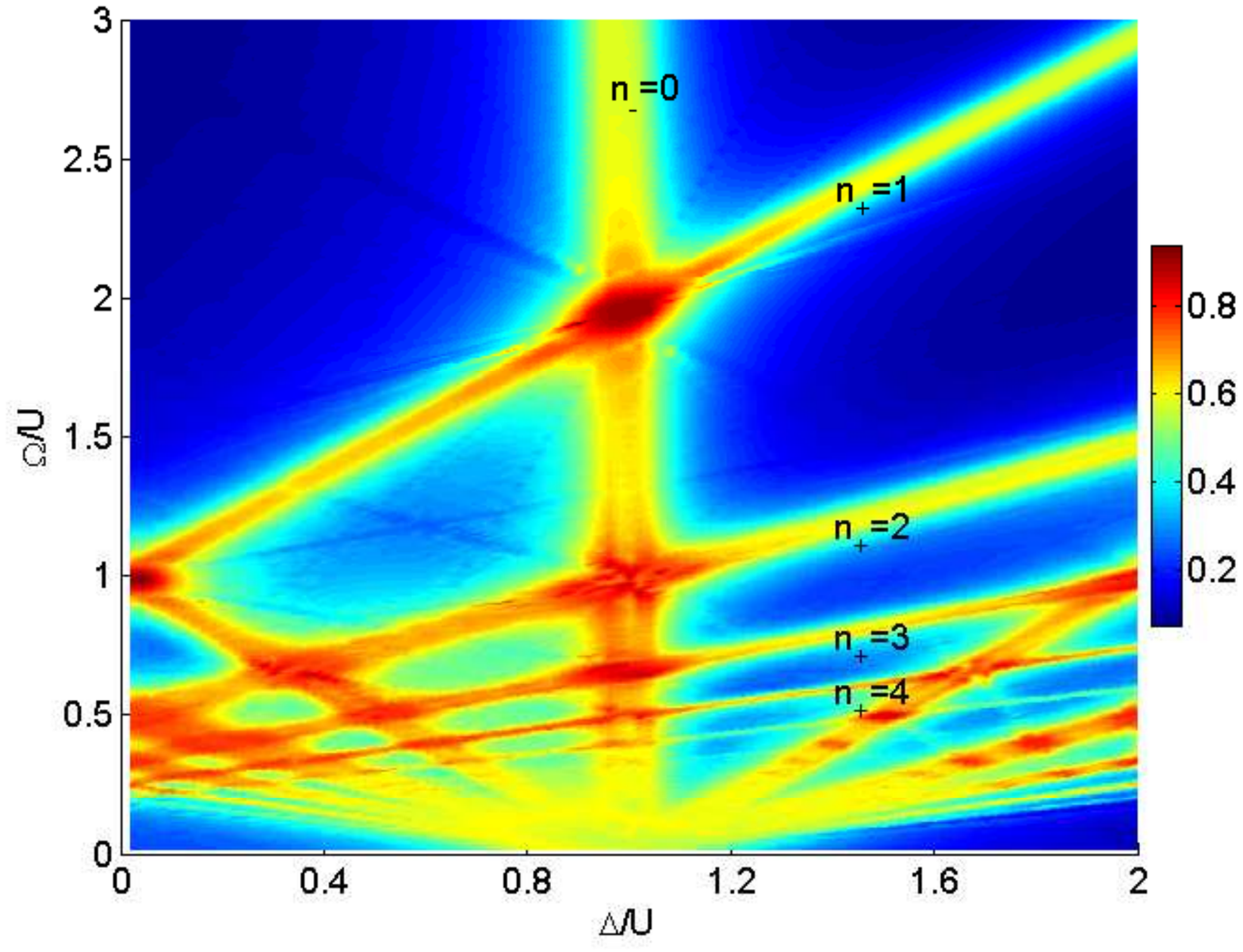}
\caption{(Color online) Topographical map of $\alpha_\textrm{max}$ versus $\Delta/U$ and $\Omega/U$ in the tilted double-well at fixed $A/\Omega=3.2$. One of the resonant branch for $n_+$ exhibit straight lines with positive slopes. The other resonant branch for $n_-$ exhibits a symmetrical fan structure from $n_-=0$ at the center to $n_-=\pm1,\pm2,\cdots$ (not specified in the figure) successively.}
\end{center}
\end{figure}

\section{summary}
In summary, we have investigated the effects of photon-assisted tunneling on the quantum coherence in a periodically drived double-well, both untitled or tilted. In the strongly interacting regime $U/\upsilon\gg 1$, the correlated tunneling coefficient is renormalized. The resonant enhancement of coherence takes place as the driving frequency matches the energy difference of the quantum transition, $n\Omega=U\pm\Delta$, which involves multiple photon absorption or emission processes.

This work is supported by the NSF of China under Grant no. 11374036 and the National 973 program under Grant no. 2012CB821403.

\end{document}